%% file: neurips_2026_vericode_workshop.tex
\documentclass{article}

\usepackage[main, final]{neurips_2026_vericode}

\usepackage[utf8]{inputenc} % allow utf-8 input
\usepackage[T1]{fontenc}    % use 8-bit T1 fonts
\usepackage{hyperref}       % hyperlinks
\usepackage{url}            % simple URL typesetting
\usepackage{booktabs}       % professional-quality tables
\usepackage{amsfonts}       % blackboard math symbols
\usepackage{nicefrac}       % compact symbols for 1/2, etc.
\usepackage{microtype}      % microtypography
\usepackage{xcolor}         % colors
\usepackage{algorithm}
\usepackage{newfloat}
\usepackage{listings}
\usepackage{graphicx} % DO NOT CHANGE THIS

\usepackage{tikz}
\usetikzlibrary{shapes.geometric, arrows, positioning}
\usepackage{booktabs}
\usepackage{amssymb}
\usepackage{longtable}
\usepackage{algpseudocode}
\usepackage{upquote}

\title{ARCHER: Agentic Rule and Compliance Harness \\ for Executable Regulations}

\author{%
  Chiraag Anand \qquad Xue Wen Tan \qquad Lionel Teo \qquad Eric Tan \\
  Infocomm Media Development Authority of Singapore \\
  \texttt{\{chiraag\_anand, tan\_xue\_wen, lionel\_teo, eric\_tan\}@imda.gov.sg} \\
}

\begin{document}

\maketitle

\begin{abstract}
\input{parts/abstract.tex}
\end{abstract}

\input{parts/introduction.tex}
\input{parts/related-work.tex}
\input{parts/methodology.tex}

\input{parts/experiments.tex}
\input{parts/limitations.tex}

\bibliographystyle{plainnat}
\bibliography{references}

\clearpage
\appendix
\input{parts/annexA.tex}
\input{parts/annexB.tex}
\input{parts/annexC.tex}

\input{parts/annexD.tex}
\input{parts/annexE.tex}
\input{parts/annexF.tex}
\input{parts/annexG.tex}

\end{document}

%% file: parts/abstract.tex
Verifying building compliance requires validating thousands of rules against large Building Information Modeling (BIM) designs, which is laborious, capital-intensive, and unscalable. Existing Automated Compliance Checkers (ACCs) are often difficult to generalize across different scenarios, as they are typically developed for highly specific rule sets and use cases. In addition, many ACCs are proprietary, meaning the underlying verification code is not released to end users, so users cannot verify whether their regulatory intent can be accurately captured. We introduce ARCHER (Agentic Rule and Compliance Harness for Executable Regulations), a test-driven, deterministically orchestrated multi-agent program-synthesis harness that generates auditable verification code from regulatory Codes of Practice, enabling transparent, adaptable, and scalable compliance checking. To characterize what makes agentic synthesis work, we evaluate a taxonomy of six harnesses of increasing agentic sophistication across four backbone models, spanning realistic data-governance tiers (from frontier third-party APIs to a fully on-premise open-weights model) on a novel dataset derived from real-world compliance scenarios. ARCHER's deterministic multi-agent orchestration achieves the highest accuracy for every backbone, improving mean union accuracy by 82\% over a na\"ive single-pass prompting baseline. Our cost--accuracy analysis further shows that using the ARCHER harness, a self-hosted open-weights model can reach 97.8\% of frontier-API accuracy at a quarter of the cost, making data-sovereign compliance checking practical.

%% file: parts/introduction.tex
\section{Introduction}

Building Information Modeling (BIM), through the Industry Foundation Classes (IFC) schema, has moved architectural design beyond conventional CAD geometry toward semantically rich, interoperable data models \citep{buildingSMART_IFC}. Yet compliance verification against these models remains largely manual and costly \citep{eastman2009automatic,aydin2022automated,madireddy2025large}, and this bottleneck grows more acute as projects scale in size and complexity \citep{wan2025automatic}. The challenge is especially pronounced in jurisdictions such as Singapore, where projects must satisfy requirements across thousands of regulatory rules \citep{BCA_BuildingControl,CORENETX_Codes}.

Codifying these requirements is difficult: natural language regulations are often ambiguous \citep{bus2018towards,hettiarachchi2025code}, while the domain experts who hold the regulatory intent typically lack the programming expertise to formalise it as executable logic \citep{eastman2009automatic,nawari2019generalized}. As the parking-lot example in Annex~\ref{annex:motivating-example} illustrates, even simple dimensional requirements can admit multiple plausible computational interpretations depending on how BIM elements are represented and measured. To address these challenges, our main contributions are as follows:
\begin{itemize}
    \item To the best of our knowledge, we are the first to apply agentic program synthesis to automated building compliance checking.
    \item We introduce \textbf{ARCHER} (\textbf{A}gentic \textbf{R}ule and \textbf{C}ompliance \textbf{H}arness for \textbf{E}xecutable \textbf{R}egulations), an effective multi-agent harness that orchestrates planning, generation, and evaluation to produce inspectable compliance checkers.
    \item We present a comprehensive cost--accuracy analysis across different models and deployment tiers, from commercial APIs to privacy-preserving on-premise setups.
    \item We release a novel benchmark dataset of real-world regulatory requirements paired with expert-labelled BIM models derived from real dataset\footnote{Code and Dataset will be released upon publication.}.
\end{itemize}

\begin{figure*}[ht]
    \centering
    \includegraphics[width=1.0\linewidth]{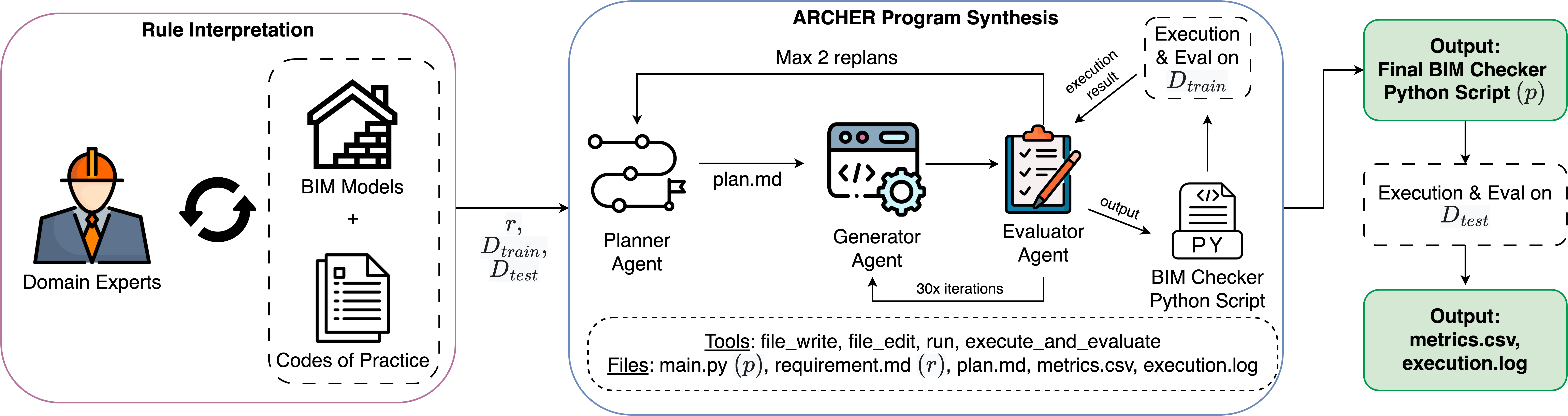}
    \caption{End-to-end architecture of ARCHER. Left: Domain experts digest regulatory Codes of Practice and BIM models to establish the Rule Interpretation ($r$) and labelled training and test pairs ($D_{train}$, $D_{test}$). Right: ARCHER's deterministic multi-agent harness drives Planner, Generator, and Evaluator agents using sandboxed tools and workspace files to iteratively synthesize, evaluate, and refine Python code (\texttt{main.py}), allowing up to 30 iterations and max 2 replans, outputting an Executable BIM Checker ($p$). The final BIM checker Python script is evaluated on the held-out test pairs ($D_{test}$).}
    \label{fig:archer-architecture}
\end{figure*}

%% file: parts/related-work.tex
\section{Related Work}

Automated Compliance Checkers (ACCs) have traditionally relied on deterministic rule-based systems or semantic reasoning approaches. Rule-based systems, including commercial BIM model-checking platforms \citep{SolibriChecking}, are effective for well-scoped checks \citep{eastman2009automatic} but are often hard-coded for a narrow set of regulatory provisions, with rigid data assumptions and proprietary implementations, making them difficult to adapt or inspect \citep{solihin2015classification,sobhkhiz2021framing}. Semantic web and ontology-based approaches \citep{pauwels2017semantic} and knowledge-graph approaches \citep{peng2023automated} improve flexibility by structuring regulatory knowledge, but fail to handle incomplete BIM metadata, complex geometry and spatial relationships \citep{xiao2026automating}.

Recent LLM-based approaches reduce some of these barriers by translating natural language regulations into computable representations \citep{fuchs2024using,chen2024automated}, generating BIM-checking scripts \citep{madireddy2025large}, or combining retrieval-augmented generation with external rule engines \citep{wan2025automatic}. However, these methods still depend heavily on predefined APIs, extracted metadata, or position-based attributes, limiting their ability to capture the spatial and contextual reasoning used in manual expert review. Broader agentic code-generation work suggests that reliability can be improved through test-driven feedback \citep{rehan2026test,liang2026scaling}, code-execution-grounded agent design \citep{wang2024executable,yang2024swe}, multi-role agents \citep{hong2024metagpt,qian2024chatdev,huang2023agentcoder}, and deterministic orchestration \citep{drammeh2025multi}.

%% file: parts/methodology.tex
\section{Our Approach: ARCHER}

To address these gaps, we introduce \textit{\textbf{ARCHER (Agentic Rule and Compliance Harness for Executable Regulations)}}, a test-driven, multi-agent program-synthesis harness that turns expert-verifiable requirement specifications into executable compliance checkers. ARCHER consumes two artefacts that domain experts can author and audit: a \textit{Rule Interpretation (RI) document} capturing the regulatory logic in plain text, and \textit{labelled BIM models} capturing its intended verdicts. It then applies \textit{Test-Driven Development (TDD)}, in which a deterministic orchestration loop drives planner, generator, and evaluator agents that iteratively generate, execute, and refine BIM Python checkers against the labelled training pair ($D_{train}$); the final generated checker is then evaluated once on the held-out test pair ($D_{test}$). The overall architecture is illustrated in Figure~\ref{fig:archer-architecture}.

\subsection{Problem Definition}
\label{sec:problem-definition}

To systematically evaluate the efficacy of agentic program synthesis for BIM compliance checking, we formalize the environment and objectives.

Let $M$ denote the space of \textbf{IFC BIM 3D models}. A 3D model $m \in M$ exposes a set of typed elements $E(m) = \{e_1, e_2, e_3, \dots\}$, each identified by a globally unique id, carrying geometry and metadata (properties, property sets, compositional relations to other elements).

A \textbf{Rule Interpretation document} $r$ translates a text-heavy Codes of Practice into an unambiguous logic flow (see Annex~\ref{annex:rule-interpretation-example}). It specifies what must hold, but not how to compute it geometrically, reflecting BIM experts' domain knowledge rather than programming expertise.

Each RI labels its relevant anchor elements with one of five \textit{\textbf{mutually exclusive outcomes}}:
\[
L = \{\mathrm{PASS}, \mathrm{FAIL}, \mathrm{ALERT}, \mathrm{MANUAL\_CHECK}, \mathrm{NA}\}.
\]
$\mathrm{PASS}$ indicates no compliance issue and $\mathrm{NA}$ indicates exemption from the rule, both requiring no further action. $\mathrm{FAIL}$ indicates a compliance issue requiring rectification of the model. $\mathrm{ALERT}$ indicates an incomplete modelling pre-condition requiring rectification, and $\mathrm{MANUAL\_CHECK}$ is reserved for scenarios that cannot be resolved by deterministic geometric checks (e.g., ambiguous shaped staircase).

For a fixed requirement $r$, a \textbf{verdict} on a model $m$ is a partial labeling of $r$'s anchor elements: a function
\begin{equation}
    y : E_y \to L, \qquad E_y \subseteq E(m),
    \label{eq:verdict}
\end{equation}
where $E_y$ denotes the elements $y$ actually labels. $y$ assigns an outcome to each anchor element $r$ deems relevant and leaves the rest of $E(m)$ unlabeled. Equivalently, $y$ is a set of pairs $\{(e, l) : e \in E_y,\, l \in L\}$ with at most one label per element. Please refer to Annex~\ref{annex:dom-example} for a worked example.

\subsection{Dataset and Test Cases}
\label{sec:dataset}

To instantiate this problem space, we construct a dataset of ten regulatory requirements (R1--R10; see Annex~\ref{annex:dataset-requirements}) derived from real Codes of Practice. For each requirement $r$, we have both labelled train and test pair:
\begin{equation}
D_{train} = (m_{train}, y^{*}_{train}), \qquad D_{test} = (m_{test}, y^{*}_{test}),
\label{eq:train-test}
\end{equation}
where $y^{*}$ is the domain expert verdict. The test pair perturbs parameters, uses disjoint building elements, and adds a few out-of-distribution variations to rigorously probe the robustness of the synthesized program. Across the ten requirements, there are a total of 245 labelled elements in $D_{train}$ and 270 labelled elements in $D_{test}$.

\subsection{Program Synthesis and Evaluation Metric}
\label{sec:program-synthesis}

Let $P$ be the space of executable Python checkers; a program $p \in P$ induces a verdict by execution against a model, $p : M \to Y$. This makes program synthesis test-driven: given $r$ and the training pair $D_{train}$, ARCHER treats $y^{*}_{train}$ as the test oracle and repeatedly generates, executes, and refines $p$ until its output on $m_{train}$ matches it (Section~\ref{sec:harness-taxonomy} details this generate--test--refine loop).

Matching cannot be reduced to comparing labels alone: an element the checker never identifies contributes no match at all. For two verdicts $y, y' \in Y(m)$ on the same model, we capture this with \textit{\textbf{union accuracy}}:
\begin{equation}
    A_{\cup}(y, y') = \frac{\bigl|\{\, e \in E_y \cap E_{y'} : y(e) = y'(e) \,\}\bigr|}{\bigl|E_y \cup E_{y'}\bigr|}.
    \label{eq:union-accuracy}
\end{equation}

$A_{\cup}$ is exactly the pass/fail signal the TDD loop optimizes against $D_{train}$; held-out generalization is then measured by re-running that same test on $D_{test}$: the synthesized program with the highest agreement to the domain expert verdict is $\hat{p}$,
\begin{equation}
    \hat{p} = \arg\max_{p \in P} A_{\cup}\bigl(p(m_{test}),\, y^{*}_{test}\bigr).
    \label{eq:synthesis-target}
\end{equation}

\subsection{Agentic Harness Architecture}
\label{sec:harness-taxonomy}

\begin{table}[t]
\centering
\scriptsize
\setlength{\tabcolsep}{4pt}
\renewcommand{\arraystretch}{1.25}
\caption{Taxonomy of harnesses evaluated, ordered by increasing sophistication of agentic structure and feedback access.}
\label{tab:harness-taxonomy}
\begin{tabular}{@{}c p{0.25\linewidth} p{0.55\linewidth}@{}}
\toprule
\textbf{ID} & \textbf{Features} & \textbf{Description} \\
\midrule
0 & Blind & A single agent must self-evaluate based on the requirement alone, with no $D_{train}$ access. \\
1 & BIM Model Aware & The agent evaluates based on execution against $m_{train}$ only. \\
2 & Full TDD & The agent evaluates against $m_{train}$ and sees metrics computed against $y^{*}_{train}$.  \newline i.e., With $D_{train}$ access. \\
3 & Planning Step & With $D_{train}$ access, single agent, but the prompt now includes an explicit planning step. \\
4 & \hbox{Multi-Agent,}\hbox{Non-Deterministic} Orchestration & With $D_{train}$ access, with planner, generator, and evaluator sub-agents provided as tools to an orchestrator agent, which decides the call order and looping. \\
5 & Multi-Agent, \hbox{Deterministic} \hbox{Orchestration} \hbox{(refer to ARCHER)} & With $D_{train}$ access, with planner, generator, and evaluator orchestrated by a fixed control loop rather than by an LLM orchestrator. \\
\bottomrule
\end{tabular}
\end{table}

\begin{algorithm}[t]
\caption{Harness 5: planner--generator--evaluator loop}
\label{alg:harness5}
\begin{algorithmic}[1]
\scriptsize
\Require{Requirement $r$; training pair $D_{train} = (m_{train}, y^{*}_{train})$; \textsc{MaxIters} $= 30$; \textsc{MaxReplans} $= 2$}
\Ensure{Synthesized checker $p$}
\State sandbox $\gets$ empty \texttt{main.py} $+$ read-only $m_{train}$
\State \textsc{replans} $\gets 0$
\State \Call{Planner}{``Write the coding plan to \texttt{plan.md} now.''} \Comment{writes \texttt{plan.md}}
\State $\textit{instr} \gets$ ``Read \texttt{plan.md} and write the first version of \texttt{main.py}.''
\For{$i \gets 1$ \textbf{to} \textsc{MaxIters}}
    \State \Call{Generator}{\textit{instr}} \Comment{writes/edits \texttt{main.py}; every write is silently scored on $m_{train}, m_{test}$ for later trajectory analysis, not used for control flow}
    \State $\textit{feedback} \gets$ \Call{Evaluator}{``Run \textsc{ExecuteAndEvaluate} and report accuracy and what to fix.''}
    \Comment{internally: $(A_\cup, \mathrm{precision}, \ldots) \gets$ \Call{ExecuteAndEvaluate}{\texttt{main.py}, $m_{train}$}; \textbf{may also} call \Call{RequestReplan}{reason} below}
    \If{$A_\cup\bigl(p(m_{train}),\, y^{*}_{train}\bigr) \geq 1.0$}
        \State \textbf{break} \Comment{early exit: perfect train agreement}
    \EndIf
    \State $\textit{instr} \gets$ ``The evaluator reported: '' $+$ \textit{feedback} $+$ `` Revise \texttt{main.py} to fix these issues.''
\EndFor
\State $p \gets$ last \texttt{main.py} written by \textsc{Generator}
\State \textbf{score} $p$ on $m_{train}$ and $m_{test}$ for reporting only (Equation~\ref{eq:synthesis-target}) \Comment{test set was never mounted during the loop}
\Return $p$
\Statex
\Function{RequestReplan}{reason} \Comment{tool exposed only to \textsc{Evaluator}; calls \textsc{Planner} directly}
    \If{\textsc{replans} $\geq$ \textsc{MaxReplans}}
        \Return ``replan limit reached -- revise \texttt{main.py} within the existing plan.''
    \EndIf
    \State \textsc{replans} $\gets$ \textsc{replans} $+ 1$
    \State \Call{Planner}{``Revise \texttt{plan.md} to address: '' $+$ reason $+$ ``; keep what already works.''}
    \Return ``\texttt{plan.md} revised -- re-read it before giving feedback.''
\EndFunction
\end{algorithmic}
\end{algorithm}

A harness $H$ drives one or more LLM agents to produce a program $p$ from a fixed $r$. Harnesses differ along two axes. The first axis is how much of $D_{train}$ the agents may consult while iterating: no $D_{train}$ access at all, $m_{train}$ alone, or full scoring feedback against $y^{*}_{train}$. The second axis is how many agents are involved and who sequences them: a single agent, a single agent with an explicit planning step, or a multi-agent pipeline coordinated by either an LLM orchestrator or a deterministic loop. We structure the program synthesis pipeline as six harnesses of increasing sophistication along both axes, each operating within a sandboxed environment that controls file access while allowing arbitrary code execution for evaluation purposes. Table~\ref{tab:harness-taxonomy} details the six harnesses evaluated in this work, ordered by increasing sophistication of agentic structure and feedback access.

\textbf{Harness 5 is our proposed configuration.} The planner first drafts a plan; the generator and evaluator then iterate for up to 30 rounds, with the evaluator executing and scoring the generated BIM python checker against $D_{train}$ each round, stopping early once union accuracy reaches 1.0. If the evaluator traces a failure to the plan rather than to a coding bug, it may trigger up to two targeted re-plans, sending control back to the planner before resuming the generate-evaluate loop. The full pseudocode of this loop is detailed in Algorithm~\ref{alg:harness5}.

%% file: parts/experiments.tex
\section{Experimental Results}
\label{sec:experiments}

We evaluate the six harnesses of Section~\ref{sec:harness-taxonomy} on the dataset of ten requirements (R1--R10, Section~\ref{sec:dataset}), each run under four backbone models (\texttt{GPT-5.5}, \texttt{GPT-5.4 Mini}, \texttt{DeepSeek-v4-flash}, and \texttt{GPT-OSS-120B-Q4KM}). Every run is scored with union accuracy $A_{\cup}$ (Equation~\ref{eq:union-accuracy}), using the best-scoring checkpoint the harness produced.

The four backbones are deliberately chosen to span the deployment spectrum an organization faces when codifying regulations over proprietary BIM data. \texttt{GPT-5.5} represents the \emph{frontier-API} tier, viable when a company may upload its designs to a third-party provider; \texttt{GPT-5.4 Mini} represents the \emph{budget-API} tier for organizations that can only afford a cheaper third-party model; \texttt{DeepSeek-v4-flash} represents an open-weights model an organization can \emph{self-host in its own cloud} when data may not leave its environment; and \texttt{GPT-OSS-120B} (4-bit \texttt{Q4KM} quantization) represents the strictest \emph{fully on-premise} tier; in our experiments it is served locally on a single NVIDIA DGX Spark. To place all four tiers on a common cost scale, we price every run at the published OpenRouter per-token rates (Table~\ref{tab:pricing}); for the self-hosted and on-premise tiers these rates act as a market proxy for serving cost, since actual amortized hardware cost depends on utilization.

\begin{table}[t]
\centering
\footnotesize
\setlength{\tabcolsep}{4pt}
\renewcommand{\arraystretch}{1.15}
\caption{Backbone models, their deployment tier, and OpenRouter list prices (USD per 1M tokens, July 2026).}
\label{tab:pricing}
\begin{tabular}{@{}l l r r@{}}
\toprule
\textbf{Model} & \textbf{Deployment tier} & \textbf{Input} & \textbf{Output} \\
\midrule
GPT-5.5 & Frontier API (third party) & \$5.00 & \$30.00 \\
GPT-5.4 Mini & Budget API (third party) & \$0.75 & \$4.50 \\
DeepSeek-v4-flash & Self-hosted (private cloud) & \$0.09 & \$0.18 \\
GPT-OSS-120B-Q4KM & On-premise (DGX Spark) & \$0.03 & \$0.18 \\
\bottomrule
\end{tabular}
\end{table}

No existing checker can serve as an external baseline on this benchmark. Commercial platforms such as Solibri require each requirement to first be hand-encoded in a proprietary rule language \citep{SolibriRulesets} and their report-style outputs do not yield the element-level verdicts needed to obtain the accuracy. Prior LLM-based checkers \citep{wan2025automatic} are incompatible with raw IFC models and are vendor-specifc. Instead, Harnesses 0--4 instantiate the field's principal design points under identical inputs, backbones, and metric: (0) single-shot generation , (1) execution-grounded generation \citep{wang2024executable}, (2--3) test-driven refinement \citep{rehan2026test}, and (4) LLM-orchestrated multi-agent pipelines \citep{hong2024metagpt}. The cross-harness comparison is thus a fair and controlled evaluation of the main architectural alternatives.

\subsection{Accuracy Across Harnesses}
\label{sec:accuracy-results}

\begin{figure*}[t]
\centering
\includegraphics[width=0.95\textwidth]{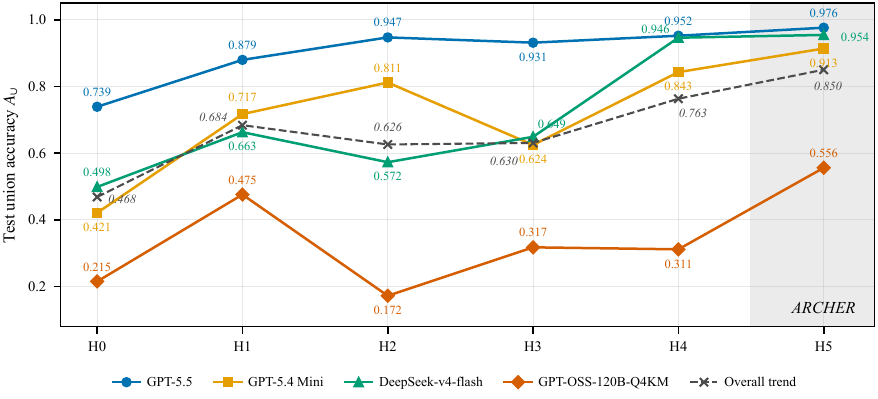}
\caption{Test union accuracy versus harness sophistication, pooled across the 10 scenarios; exact values are annotated at each point and the dashed line shows the overall trend (mean over the four backbones). ARCHER (Harness 5) is the best configuration for every backbone model, lifting the self-hosted \texttt{DeepSeek-v4-flash} to within 2.2 points of the frontier tier.}
\label{fig:accuracy-vs-harness}
\end{figure*}

Figure~\ref{fig:accuracy-vs-harness} presents the union accuracy achieved by each harness and backbone model, with exact values annotated at each point; the dashed grey line traces the overall trend (mean across the four backbones).

Three findings stand out. \textbf{(1) ARCHER configuration is uniformly best.} Harness 5 achieves the highest union accuracy for every backbone, raising the overall mean from 0.4681 (Harness 0, blind baseline) to 0.8498, a gain of 38.2 accuracy points, or an 82\% relative improvement. \textbf{(2) Deterministic orchestration beats LLM orchestration.} Harness 5 improves over the LLM-orchestrated Harness 4 for all four models (by $+$0.008 to $+$0.245), confirming that a fixed planner--generator--evaluator control loop is more reliable than delegating call order to an orchestrator agent, particularly for weaker backbones, which struggle to sequence their own sub-agents. \textbf{(3) Feedback without structure hurts weaker models.} Exposing full TDD feedback in a single-agent prompt (Harness 2) helps the strong API models (GPT-5.5 rises to 0.9468) but \emph{degrades} the open-weights models relative to Harness 1 (\texttt{DeepSeek-v4-flash} drops from 0.6625 to 0.5725; \texttt{GPT-OSS-120B-Q4KM} collapses from 0.4754 to 0.1719): weaker models are unable to convert dense scoring feedback into targeted code revisions unless the harness decomposes that work across the three specialized agents. The structured multi-agent harnesses (4--5) recover and surpass these losses, with \texttt{DeepSeek-v4-flash} jumping from 0.6486 (Harness 3) to 0.9459 (Harness 4) and 0.9542 (Harness 5).

\subsection{Token Consumption Analysis}
\label{sec:token-consumption}

\begin{figure*}[t]
\centering
\includegraphics[width=0.95\linewidth]{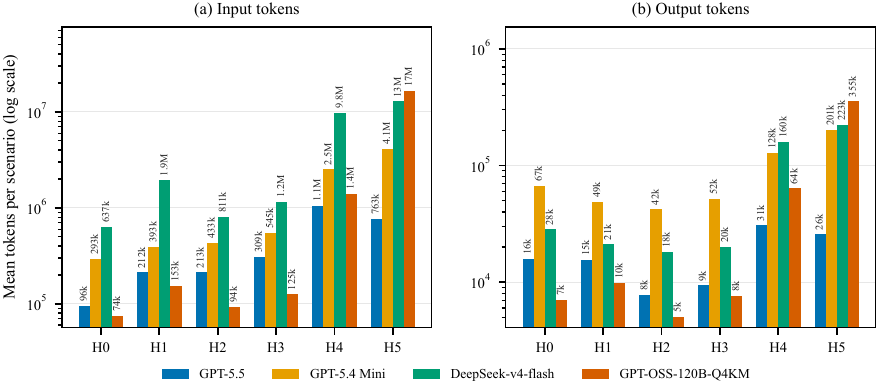}
\caption{Mean token consumption across scenarios by harness configuration and backbone model (log scale). Single-agent harnesses (0--3) stay within a narrow band; multi-agent harnesses (4--5) grow by an order of magnitude, and the growth is steepest for the weakest backbone.}
\label{fig:token-consumption}
\end{figure*}

Figure~\ref{fig:token-consumption} reports the mean input and output token consumption across scenarios by harness configuration and backbone model, with exact counts annotated on each bar.

As expected, token usage scales with harness complexity: the single-agent harnesses (0--3) remain within a comparatively narrow band, while the multi-agent architectures (4--5) consume an order of magnitude more tokens due to iterative planner--generator--evaluator loops and execution-trace evaluations.

More interesting is how consumption scales \emph{inversely with model capability}. Moving from Harness 0 to Harness 5 multiplies input tokens by only $8.0\times$ for \texttt{GPT-5.5}, but by $14.1\times$ for \texttt{GPT-5.4 Mini}, $20.2\times$ for \texttt{DeepSeek-v4-flash}, and $223.7\times$ for \texttt{GPT-OSS-120B-Q4KM}. Because ARCHER's evaluator re-invokes the generator until train accuracy converges (or the iteration cap is hit), weaker models pay for their mistakes in tokens: \texttt{GPT-OSS-120B-Q4KM} consumes $21.8\times$ the input tokens of \texttt{GPT-5.5} under Harness 5 while reaching barely half its accuracy. In addition, \texttt{GPT-5.5} is the only backbone whose input consumption \emph{drops} from Harness 4 to Harness 5 (a mean of 1,053,584 to 762,804 tokens per scenario): once a capable model is placed in a deterministic loop, the undeterministic orchestration chatter of Harness 4 disappears and early stopping triggers sooner. Output tokens follow the same pattern but remain a small fraction of input volume (2--5\% at Harness 5 across models), so total cost at high harness levels is dominated by repeated context re-reading rather than code generation itself.

\subsection{Cost--Accuracy Trade-off}
\label{sec:cost-analysis}

\begin{figure}[!b]
\centering
\includegraphics[width=0.7\linewidth]{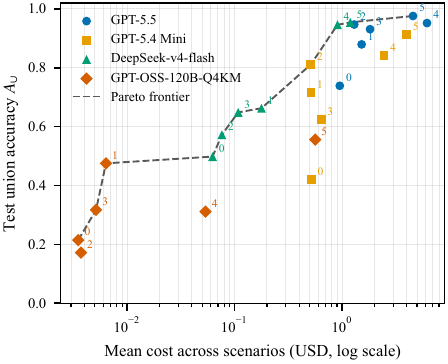}
\caption{Cost--accuracy trade-off across all 24 (harness, model) configurations. Each marker is one run of a backbone model under one harness; the small number beside each marker is the harness ID (0--5, cf.\ Table~\ref{tab:harness-taxonomy}), so, e.g., a green triangle labeled ``5'' is \texttt{DeepSeek-v4-flash} under ARCHER (Harness 5). The dashed line traces the Pareto frontier: configurations for which no alternative is both cheaper and more accurate. The frontier is populated almost entirely by the self-hostable models at low-to-mid budgets and by \texttt{DeepSeek-v4-flash} and \texttt{GPT-5.5} under ARCHER at the high end.}
\label{fig:cost-pareto}
\end{figure}

Combining token counts with per-token prices yields the monetary cost of synthesizing a checker. Table~\ref{tab:cost} reports the average cost across scenarios by harness configuration, and backbone model. Figure~\ref{fig:cost-pareto} plots every (harness, model) configuration in the cost--accuracy plane together with its Pareto frontier.

\begin{table}[t]
\centering
\footnotesize
\setlength{\tabcolsep}{4pt}
\renewcommand{\arraystretch}{1.15}
\caption{Mean synthesis cost (USD) across scenarios by harness configuration and backbone model, at the rates of Table~\ref{tab:pricing}.}
\label{tab:cost}
\begin{tabular}{@{}c rrrr@{}}
\toprule
\textbf{Harness} & \textbf{GPT-5.5} & \textbf{GPT-5.4 Mini} & \textbf{DeepSeek V4 Flash} & \textbf{GPT-OSS} \\
\midrule
0 & 0.9546 & 0.5204 & 0.0624 & 0.0035 \\
1 & 1.5245 & 0.5144 & 0.1782 & 0.0064 \\
2 & 1.3007 & 0.5140 & 0.0763 & 0.0037 \\
3 & 1.8264 & 0.6414 & 0.1081 & 0.0051 \\
4 & 6.1904 & 2.4603 & 0.9080 & 0.0537 \\
5 & 4.5860 & 4.0017 & 1.1969 & 0.5637 \\
\bottomrule
\end{tabular}
\end{table}

Even the most expensive configuration (\texttt{GPT-5.5} under Harness 4) costs \$6.19 per requirement on average, negligible against the engineering hours it replaces; the economics of agentic codification are therefore favorable across the board. The more consequential question is \emph{which tier an organization should operate}, and here three results carry practical weight.

\textbf{Harness sophistication is a better investment than model price.} The Pareto frontier in Figure~\ref{fig:cost-pareto} ascends primarily by increasing harness level within a model family rather than by switching to a pricier backbone. For every backbone, ARCHER's accuracy gain over the blind baseline costs at most a few dollars per requirement: the harness converts a modest token budget into 24--49 points of accuracy, with the return per dollar largest exactly where per-token prices are lowest (moving from Harness 0 to 5 buys 3.8 accuracy points per dollar spent on the full 10-scenario run for \texttt{DeepSeek-v4-flash} and 6.0 for \texttt{GPT-OSS-120B-Q4KM}, versus 0.5 points per dollar for \texttt{GPT-5.5}).

\textbf{A cheap per-token price does not guarantee a cheap task.} \texttt{GPT-5.4 Mini} is priced $6.7\times$ below \texttt{GPT-5.5} per input token, yet under ARCHER it consumes $5.4\times$ the input and $7.8\times$ the output tokens, so its mean per-scenario cost (\$4.00) nearly matches \texttt{GPT-5.5}'s (\$4.59) while its accuracy is 6.2 points lower. Indeed, under ARCHER \texttt{GPT-5.4 Mini} falls off the Pareto frontier entirely: the self-hosted \texttt{DeepSeek-v4-flash} delivers 4.1 more accuracy points at $3.3\times$ lower cost. Task-level cost, the price multiplied by the tokens a model actually needs to converge, is the metric that matters when provisioning an agentic pipeline.

\textbf{Data-sovereign deployments are no longer a large accuracy sacrifice.} Comparing each tier's best operating point (all of which occur under ARCHER, Harness 5), an organization that can share data with a frontier provider obtains 0.9759 at \$4.59 per scenario on average. An organization restricted to its own cloud reaches 0.9542 with self-hosted \texttt{DeepSeek-v4-flash} (97.8\% of frontier accuracy at roughly a quarter of the cost), which we attribute directly to ARCHER's deterministic orchestration, since the same model languishes at 0.50--0.66 under the single-agent harnesses. Only the strictest tier still pays a real penalty: the 4-bit \texttt{GPT-OSS-120B} served on a single DGX Spark peaks at 0.5557. Even so, ARCHER improves the model's blind-baseline accuracy (0.2148 to 0.5557) by more than double and delivers the highest accuracy-per-dollar of any configuration, suggesting that fully air-gapped compliance checking becomes viable as on-premise-servable open-weights models continue to be better: the harness is already in place to exploit them.

The aggregate view above masks substantial variation across the ten requirements: some scenarios are intrinsically harder, consume disproportionate token budgets, and reward escalation to a stronger backbone, while for the majority the self-hostable tiers already match the frontier model. Readers interested in results for specific scenarios (including a per-requirement difficulty ranking, per-requirement synthesis costs, and an analysis of when a more expensive model is worth it) are referred to Annex~\ref{annex:per-scenario-analysis}.

All figures and derived statistics in this section are reproducible from the scripts and datasets accompanying the paper.

%% file: parts/limitations.tex
\section{Limitations and Future Work}
ARCHER is best understood as a human-agent collaborative system rather than a fully autonomous one. Before synthesis begins, domain experts must author the RI document (Annex~\ref{annex:rule-interpretation-example}) and label the training and test models, upfront effort that our cost analysis does not account for; the harness assumes these artefacts faithfully capture regulatory intent, and how reliably experts author and audit them in practice remains to be studied. Auditability is correspondingly layered: synthesized checkers run to hundreds of lines of geometry-heavy Python (Annex~\ref{annex:r3-compliance-code}), so non-programming experts audit the system through its expert-facing artefacts (the RI logic, labelled verdicts, and per-element outputs) while code-level inspection requires an engineer, though unlike proprietary checkers the code is at least available to inspect. Beyond this, ARCHER treats each rule in isolation, without modelling dependencies or contradictions across rules; sub-1.0 accuracy shortfalls are not always easy to attribute to the requirement specification, the test cases, or a coding bug; and executable checkers remain bounded by the limits of BIM schemas and computational geometry. Future work could address these gaps with semantic knowledge graphs for rule dependencies, auto-corrective post-processing, modular intermediate representations to localise errors, and test-time methods such as spatial reasoning networks or vision-language models for cases outside deterministic geometric checking.

%% file: parts/annexA.tex
\section{Motivating Example: Ambiguity in Geometric Rule Operationalisation}
\label{annex:motivating-example}

Consider a simple regulation: ``each parking lot shall be 2.4~m wide and 4.8~m long.'' While the rule specifies what must hold, it does not specify how the governing dimensions should be measured. This omission can lead to mutually incompatible implementations. The same parking bay may be authored as a filled solid, a set of 2D boundary line markings, or an object accompanied by tool-exported or modeller-supplied width and length properties. Each representation admits a different, individually reasonable measurement strategy, but these strategies may disagree in cases that determine compliance.

Table~\ref{tab:parking-measurement-methods} illustrates four candidate measurement methods. Each method is valid in some scenarios but fails in others, highlighting why compliance checking cannot rely solely on metadata, positional points, or a single fixed geometric abstraction.

\begin{table*}[htbp]
\centering
\footnotesize
\setlength{\tabcolsep}{3pt}
\renewcommand{\arraystretch}{1.15}
\caption{Candidate methods for checking parking-lot dimensions under different BIM representations.}
\label{tab:parking-measurement-methods}
\begin{tabular}{p{0.38\linewidth}cccc}
\toprule
\textbf{Input class / representation}
& \textbf{\shortstack{Property\\only}}
& \textbf{\shortstack{Property +\\geometry fallback}}
& \textbf{\shortstack{2D\\footprint}}
& \textbf{\shortstack{Min-area\\bounding box}} \\
\midrule
Rectangular bay with valid property
& \(\checkmark\) & \(\checkmark\) & \(\checkmark\) & \(\checkmark\) \\

Property present but stale
& \(\times\) & \(\times\) & \(\checkmark\) & \(\checkmark\) \\

Line-marking bay with no enclosed area
& \(\times\) & \(\times\) & \(\times\) & \(\checkmark\) \\

Non-rectangular or skewed bay
& \(\times\) & \(\checkmark\) & \(\checkmark\) & \(\times\) \\

Clear width with column encroachment
& \(\times\) & \(\times\) & \(\times\) & \(\times\) \\
\bottomrule
\end{tabular}
\end{table*}

One might expect that making the rule more precise would resolve this ambiguity. However, doing so introduces a different problem: specifying a single computational method requires geometric and computational vocabulary. For example, the rule may need to define whether dimensions should be measured using an axis-aligned bounding box, a minimum-area bounding box, or a 2D footprint obtained by projecting and unioning mesh faces onto the world (x)-(y) plane. Such formulations are often opaque to BIM domain experts, even though these experts are responsible for determining which interpretation best reflects regulatory intent and real-world practice.

This creates a separation between authority over intent and authority over implementation. The expert who understands the meaning of the rule may not be able to inspect how it is computed, while the developer who implements the computation may not be best placed to judge whether it faithfully captures the rule. This separation makes compliance logic harder to validate, less transparent to end users, and more difficult to update when regulations change.

%% file: parts/annexB.tex
\section{Bounds of Computational Geometry}
\label{annex:geometry-bounds}

\begin{figure}[htbp]
    \centering
    \includegraphics[width=0.8\linewidth]{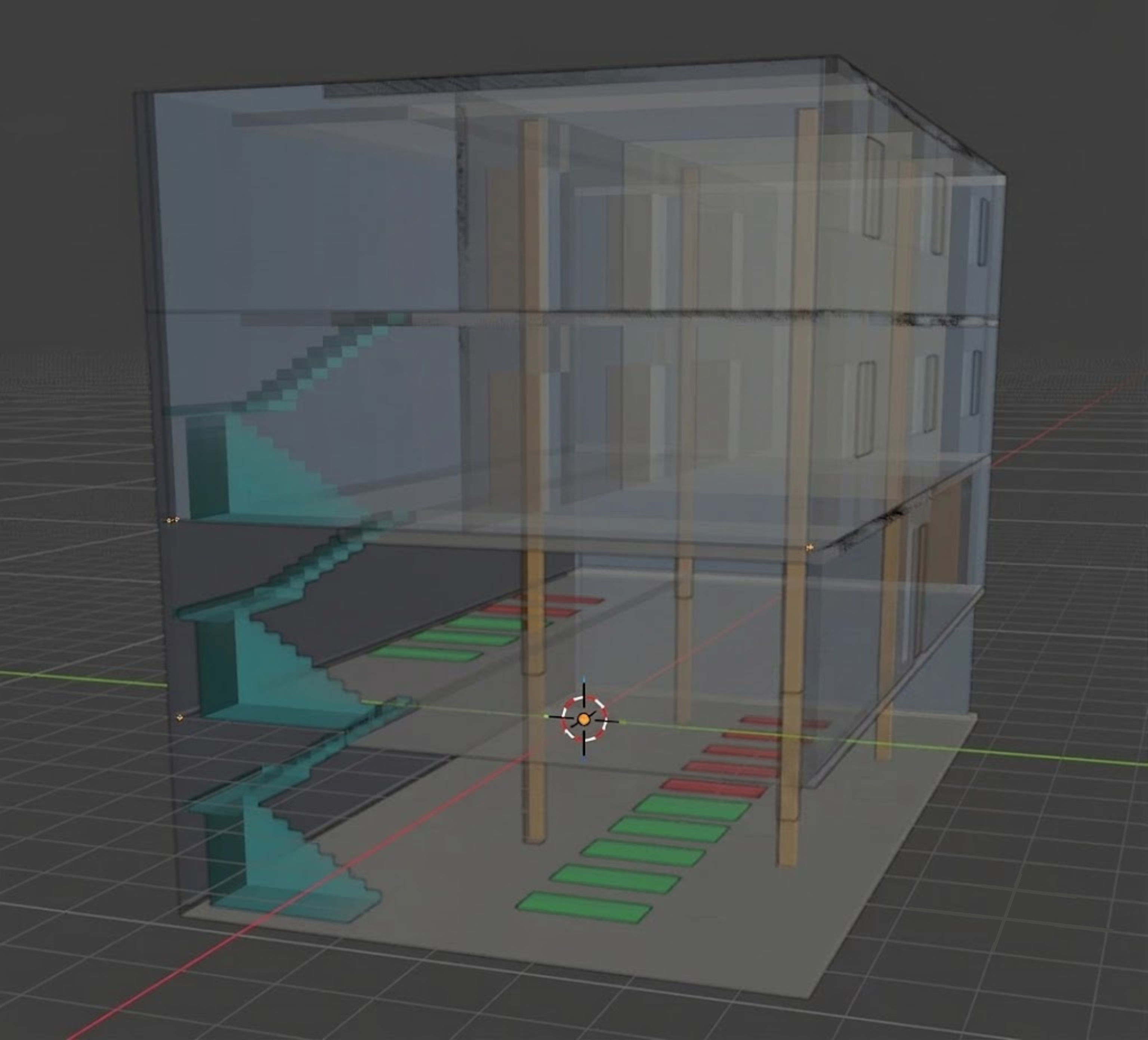}
    \caption{A rendered multi-storey BIM model with staircases, columns, doors, and window openings. Even a single object category such as a staircase varies in form and placement across storeys.}
    \label{fig:bim-geometry-variety}
\end{figure}

Annex~\ref{annex:motivating-example} showed, by example, that no single geometric method is correct across every BIM representation of a simple dimensional rule. This is not specific to the parking-lot example. It is a general limit of any executable checker $p \in P$ (Section~\ref{sec:program-synthesis}). A checker implements a fixed, finite composition of geometric primitives (bounding boxes, projections, distance thresholds, containment and adjacency tests) evaluated over a local neighbourhood of each element $e \in E(m)$. Regulatory intent, on the other hand, refers to a functional property of the space, such as whether a person can pass through unobstructed, that a human inspects holistically and that stays the same across many different ways of authoring the same design. The space of IFC models $M$ admits unboundedly many structurally distinct ways of modelling an equivalent physical layout, so a fixed procedure's domain of correctness is always a strict subset of $M$. Extending the procedure to cover one new layout just moves the boundary rather than removing it. This is why the label space $L$ (Section~\ref{sec:problem-definition}) includes $\mathrm{MANUAL\_CHECK}$. Figure~\ref{fig:bim-geometry-variety} shows an example of a BIM model: a multi-storey building containing staircases, columns, doors, and window openings.

These bounds motivate two design choices in ARCHER (Section~\ref{sec:harness-taxonomy}). First, $\mathrm{MANUAL\_CHECK}$ is treated as a first-class, legitimate output rather than something to engineer away. Second, the synthesized checker is kept as transparent, editable Python rather than an opaque proprietary engine, so that when a configuration outside the checker's geometric assumptions turns up, a domain expert can see exactly which assumption failed and decide whether to accept the manual escalation or extend the checker's geometry for that configuration.

%% file: parts/annexC.tex
\section{Example of Rule Interpretation}
\label{annex:rule-interpretation-example}

\begin{figure*}[htbp]
    \centering
    \includegraphics[width=0.8\linewidth]{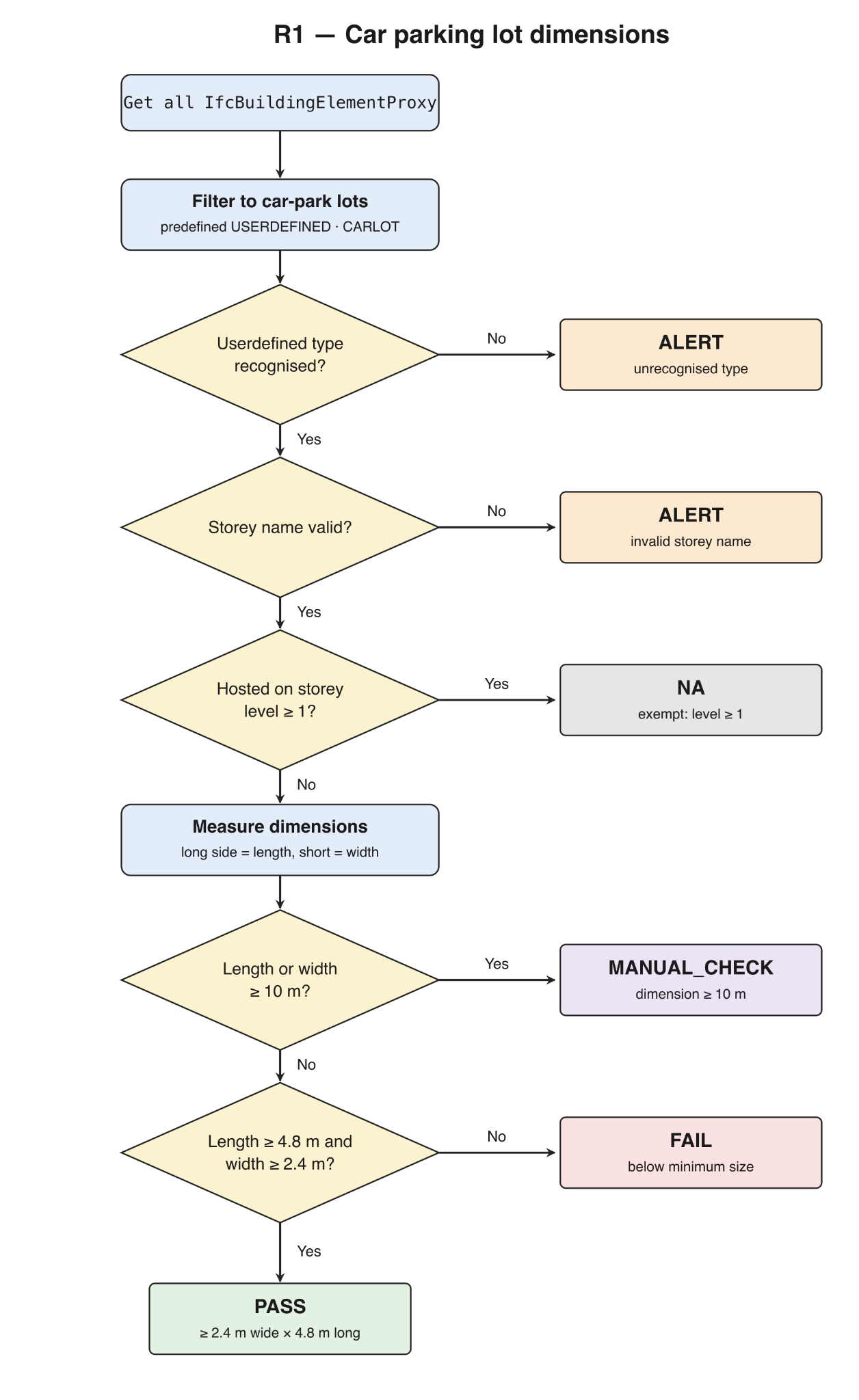}
    \caption{Rule Interpretation for R1 (car parking lot dimensions), drawn as a flow chart to show the logical branches the synthesized checker must implement.}
    \label{fig:rule-interpretation-example}
\end{figure*}

Section~\ref{sec:problem-definition} defines a Rule Interpretation document $r \in R$ as a translation of a Codes of Practice provision into an unambiguous logic flow, one that fixes what must hold while leaving how to compute it geometrically to the synthesis harness (Section~\ref{sec:harness-taxonomy}). Figure~\ref{fig:rule-interpretation-example} shows a concrete $r$ for requirement R1, the same car-parking-lot dimension rule used as the motivating example in Annex~\ref{annex:motivating-example}: each lot must be at least 2.4~m wide and 4.8~m long.

The flow chart walks through every element that could plausibly be a car parking lot and assigns it one of the five labels in $L$ (Section~\ref{sec:problem-definition}) before any dimension is measured. An element with an unrecognised or invalid type, or hosted on a storey outside the valid naming scheme, returns $\mathrm{ALERT}$: these are modelling errors, not compliance failures, and the fix is to correct the BIM model rather than the checker. A lot on storey level 1 or above is out of scope for this rule and returns $\mathrm{NA}$. A lot whose length or width is at least 10~m is flagged as suspiciously large and returns $\mathrm{MANUAL\_CHECK}$, since a car parking lot of that size is more likely to be a modelling or classification error than a genuine oversized lot, and deciding which is exactly the kind of judgement call the label space reserves for a human. Only after these checks does the flow reach the actual dimension test, returning $\mathrm{PASS}$ or $\mathrm{FAIL}$ depending on whether the measured length and width meet the 2.4~m by 4.8~m threshold.

Note what the RI document does not do: it does not say how "length" and "width" are to be measured from the underlying geometry. That question, and the ambiguity it introduces across differently authored BIM representations, is the subject of Annex~\ref{annex:motivating-example}. The RI document's job is to fix intent, exemptions, and escalation points; resolving the remaining geometric ambiguity is left to the synthesis harness.

%% file: parts/annexD.tex
\section{Worked Example: The Labeled Elements of a Verdict}
\label{annex:dom-example}

To make the $E_y$ notation from Section~\ref{sec:problem-definition} concrete, consider a small model $m$ exposing five elements, $E(m) = \{e_1, e_2, e_3, e_4, e_5\}$, $e_1$--$e_4$ are parking bays and $e_5$ is a column encroaching into bay $e_4$'s clear width.

Suppose the requirement $r$ only concerns bays, so its anchor elements are $\{e_1, e_2, e_3, e_4\}$; the column $e_5$ is metadata a checker may consult, but it is never itself given a verdict. A checker $p$ evaluates each bay in turn:
\begin{itemize}
    \item $e_1$, $e_2$: clear width comfortably exceeds the threshold $\to \mathrm{PASS}$.
    \item $e_3$: clear width is below the threshold $\to \mathrm{FAIL}$.
    \item $e_4$: the column encroachment defeats the checker's bounding-box measurement, so it cannot compute a clear width at all: it is left \textbf{out of the verdict entirely}, not assigned any label.
\end{itemize}
The resulting verdict is $p(m) = \{(e_1, \mathrm{PASS}),$ $(e_2, \mathrm{PASS}),$ $(e_3, \mathrm{FAIL})\}$, so
\[
    E_{p(m)} = \{e_1, e_2, e_3\} \subsetneq E(m).
\]

Two different reasons keep an element out of $E_{p(m)}$, and it is worth distinguishing them: $e_5$ is absent because it is not an anchor element of $r$ at all, while $e_4$ is absent despite being one, because the checker failed to resolve an outcome for it. Note that $\mathrm{MANUAL\_CHECK}$ would \emph{not} have caused this: it is a definite label, just like $\mathrm{PASS}$ or $\mathrm{FAIL}$, that defers the decision to a person rather than a missing one. Had the checker returned $(e_4, \mathrm{MANUAL\_CHECK})$, $e_4$ would still be in $E_{p(m)}$. Only a genuinely missing entry (an element the checker never identified or could not evaluate at all) removes it from $E_{p(m)}$.

This distinction is what union accuracy (Equation~\ref{eq:union-accuracy}) scores. If the expert resolves all four bays, $y^{*}_{test} = \{(e_1, \mathrm{PASS}), (e_2, \mathrm{PASS}), (e_3, \mathrm{FAIL}), (e_4, \mathrm{FAIL})\}$, then $E_{y^{*}_{test}} = \{e_1, e_2, e_3, e_4\}$. The checker agrees with the expert everywhere it produced a label, yet $e_4 \in E_{y^{*}_{test}} \setminus E_{p(m)}$ still enlarges the denominator $|E_{p(m)} \cup E_{y^{*}_{test}}|$ without being able to contribute a match in the numerator: a checker is penalized for silently missing an element, not just for mislabeling one.

\paragraph{How $r$, $m$, and $y$ relate.} The example above fixes one of each. Table~\ref{tab:r-m-y} separates what each one controls.

Changing $r$ while holding $m$ fixed can change which elements are even eligible to appear in $E_y$ (a different rule may treat a different subset of the same five elements as anchor elements); changing $m$ while holding $r$ fixed changes the concrete elements being labeled under the same criteria. Because a given evaluation always fixes $r$ first and varies $m$ (as in $D_{train}$ and $D_{test}$, Equation~\ref{eq:train-test}), we suppress $r$ from the notation $y$, $Y(m)$, and $E_y$, but every one of them is implicitly relative to whichever $r$ is under discussion.

\begin{table*}[htbp]
\centering
\footnotesize
\setlength{\tabcolsep}{4pt}
\renewcommand{\arraystretch}{1.3}
\caption{The three axes of the problem.}
\label{tab:r-m-y}
\begin{tabular}{p{0.06\linewidth}p{0.13\linewidth}p{0.36\linewidth}p{0.35\linewidth}}
\toprule
\textbf{Symbol} & \textbf{Role} & \textbf{Fixes / controls} & \textbf{In the example above} \\
\midrule
$m$ & model & which elements exist at all, $E(m)$ & $E(m) = \{e_1, \dots, e_5\}$ \\
$r$ & requirement & which elements of $E(m)$ are anchor elements, and what $\mathrm{PASS}$/$\mathrm{FAIL}$/etc.\ mean for them & anchor elements $= \{e_1, e_2, e_3, e_4\}$; $e_5$ is excluded \\
$y$ & verdict & a labeling of $m$'s elements under $r$'s criteria; only meaningful for one specific $(r, m)$ pair, never on its own & $p(m)$ and $y^{*}_{test}$, e.g.\ $E_{p(m)} = \{e_1, e_2, e_3\}$ \\
\bottomrule
\end{tabular}
\end{table*}

%% file: parts/annexE.tex
\onecolumn
\section{Generated Compliance Code for Motorcycle Lot Dimension Criteria (R3)}
\label{annex:r3-compliance-code}

The complete Python compliance code synthesized by ARCHER for requirement R3 (Motorcycle Lot Dimension Criteria) is listed below:

\begin{lstlisting}[language=Python]
import ifcopenshell
import trimesh
import shapely
import numpy as np
from models import OutputCheck
import ifcopenshell.geom
from ifcopenshell.util.element import get_psets
import re
from shapely.geometry import Polygon
from shapely.ops import unary_union

REQUIREMENT_ID = "R3"
REQUIREMENT_TEXT = "Smith Road Site Motorcycle Lot Dimension Criteria"
KNOWN_USER_DEFINED_TYPES = {
    "carlot",
    "bicyclelot",
    "motorcyclelot",
    "shaft",
    "manhole",
    "tree",
    "planterbox",
    "egressindicatorbox",
}
ALLOWED_BAY_USAGE = {"STANDARD", "TANDEM", "EV", "LOADING"}
ALLOWED_PARKING_ENVIRONMENT = {"OPEN", "COVERED", "SEMI"}
DIM_TOL = 0.005
CONTAIN_2D_TOL = 0.02
VERT_TOL = 0.05
TARGET_ADDRESS = "123 smith road"
LEVEL_RE = re.compile(r"^level\s+([1-9]\d*)$", re.IGNORECASE)
BASEMENT_RE = re.compile(r"^basement\s+([1-9]\d*)$", re.IGNORECASE)

def _norm_text(value):
    if value is None:
        return None
    text = re.sub(r"\s+", " ", str(value)).strip()
    return text if text else None

def _norm_lower(value):
    text = _norm_text(value)
    return text.lower() if text is not None else None

def _norm_upper(value):
    text = _norm_text(value)
    return text.upper() if text is not None else None

def _make_result(element, result, reason):
    return OutputCheck(
        requirement_id=REQUIREMENT_ID,
        requirement=REQUIREMENT_TEXT,
        element_guid=element.GlobalId,
        result=result,
        reason=reason,
    )

def _get_property_text(element, pset_name, prop_name, normalizer=_norm_text):
    try:
        psets = get_psets(element)
    except Exception:
        return None
    pset = psets.get(pset_name)
    if not isinstance(pset, dict):
        return None
    value = pset.get(prop_name)
    if value is None:
        return None
    return normalizer(value)

def _extract_building_address_text(building):
    addr = getattr(building, "BuildingAddress", None)
    if not addr:
        return None
    if isinstance(addr, str):
        return _norm_text(addr)

    parts = []
    address_lines = getattr(addr, "AddressLines", None)
    if address_lines:
        for line in address_lines:
            text = _norm_text(line)
            if text:
                parts.append(text)
    if not parts:
        for attr in ("InternalLocation", "Description", "Town", "Region", "PostalCode", "Country"):
            text = _norm_text(getattr(addr, attr, None))
            if text:
                parts.append(text)
    if not parts:
        return None
    return _norm_text(", ".join(parts))

def _resolve_building_and_storey(element):
    start_nodes = []
    for rel in getattr(element, "ContainedInStructure", []) or []:
        spatial = getattr(rel, "RelatingStructure", None)
        if spatial is not None:
            start_nodes.append(spatial)
    for rel in getattr(element, "Decomposes", []) or []:
        parent = getattr(rel, "RelatingObject", None)
        if parent is not None:
            start_nodes.append(parent)

    visited = set()
    stack = list(start_nodes)
    buildings = {}
    storeys = {}

    while stack:
        current = stack.pop()
        if current is None:
            continue
        current_id = current.id()
        if current_id in visited:
            continue
        visited.add(current_id)

        if current.is_a("IfcBuilding"):
            buildings[current.id()] = current
        if current.is_a("IfcBuildingStorey"):
            storeys[current.id()] = current

        for rel in getattr(current, "Decomposes", []) or []:
            parent = getattr(rel, "RelatingObject", None)
            if parent is not None:
                stack.append(parent)

    building = list(buildings.values())[0] if len(buildings) == 1 else None
    storey = list(storeys.values())[0] if len(storeys) == 1 else None
    return building, storey

def _geometry_cache_entry(element, settings, cache):
    key = element.id()
    if key in cache:
        return cache[key]

    try:
        shape = ifcopenshell.geom.create_shape(settings, element)
        verts_flat = np.array(shape.geometry.verts, dtype=float)
        faces_flat = np.array(shape.geometry.faces, dtype=int)
        if verts_flat.size == 0 or faces_flat.size == 0:
            cache[key] = {"ok": False, "reason": "empty mesh"}
            return cache[key]
        verts = verts_flat.reshape((-1, 3))
        faces = faces_flat.reshape((-1, 3))
    except Exception as exc:
        cache[key] = {"ok": False, "reason": f"geometry generation failed: {exc}"}
        return cache[key]

    triangles = []
    for face in faces:
        tri_xy = verts[face][:, :2]
        try:
            poly = Polygon(tri_xy)
        except Exception:
            continue
        if poly.is_empty or poly.area <= 1e-10:
            continue
        triangles.append(poly)

    if not triangles:
        cache[key] = {"ok": False, "reason": "degenerate projected footprint"}
        return cache[key]

    try:
        footprint = unary_union(triangles)
    except Exception as exc:
        cache[key] = {"ok": False, "reason": f"footprint union failed: {exc}"}
        return cache[key]

    if footprint.is_empty:
        cache[key] = {"ok": False, "reason": "empty footprint"}
        return cache[key]
    if not footprint.is_valid:
        try:
            footprint = footprint.buffer(0)
        except Exception as exc:
            cache[key] = {"ok": False, "reason": f"footprint repair failed: {exc}"}
            return cache[key]
    if footprint.is_empty:
        cache[key] = {"ok": False, "reason": "empty footprint after repair"}
        return cache[key]

    cache[key] = {
        "ok": True,
        "verts": verts,
        "faces": faces,
        "min_z": float(np.min(verts[:, 2])),
        "max_z": float(np.max(verts[:, 2])),
        "footprint": footprint,
    }
    return cache[key]

def _measure_length_width(footprint):
    candidate = footprint
    if candidate.is_empty:
        return None
    if candidate.area <= 1e-10:
        candidate = candidate.convex_hull
    mrr = candidate.minimum_rotated_rectangle
    if mrr.is_empty or not hasattr(mrr, "exterior"):
        return None
    coords = list(mrr.exterior.coords)
    if len(coords) < 5:
        return None
    lengths = []
    for i in range(4):
        p1 = np.array(coords[i][:2], dtype=float)
        p2 = np.array(coords[i + 1][:2], dtype=float)
        lengths.append(float(np.linalg.norm(p2 - p1)))
    positive = [x for x in lengths if x > 1e-8]
    if len(positive) < 2:
        return None
    length = max(positive)
    width = min(positive)
    return length, width

def _storey_classification(storey):
    name = _norm_lower(getattr(storey, "Name", None))
    if name is None:
        return None, None
    if LEVEL_RE.match(name):
        return "level", name
    if BASEMENT_RE.match(name):
        return "basement", name
    return None, name

def main(ifc_filepath) -> list[OutputCheck]:
    model = ifcopenshell.open(ifc_filepath)
    results = []
    # compliance code
    settings = ifcopenshell.geom.settings()
    settings.set(settings.USE_WORLD_COORDS, True)
    settings.set(settings.DISABLE_OPENING_SUBTRACTIONS, True)

    geometry_cache = {}
    userdefined_proxies = []
    for proxy in model.by_type("IfcBuildingElementProxy"):
        if _norm_upper(getattr(proxy, "PredefinedType", None)) == "USERDEFINED":
            userdefined_proxies.append(proxy)

    spaces_by_storey = {}
    for space in model.by_type("IfcSpace"):
        _, storey = _resolve_building_and_storey(space)
        if storey is not None:
            spaces_by_storey.setdefault(storey.id(), []).append(space)

    for proxy in userdefined_proxies:
        building, storey = _resolve_building_and_storey(proxy)
        if building is None:
            results.append(_make_result(proxy, "ALERT", "Hosting building could not be resolved from spatial structure."))
            continue

        address_text = _extract_building_address_text(building)
        if address_text is None:
            results.append(_make_result(proxy, "ALERT", "Hosting building BuildingAddress is missing or unreadable."))
            continue

        normalized_address = _norm_lower(address_text)
        if normalized_address != TARGET_ADDRESS:
            results.append(_make_result(proxy, "NA", f"Hosting building address is '{address_text}', not 123 Smith Road."))
            continue

        object_type = _norm_lower(getattr(proxy, "ObjectType", None))
        if object_type == "motorcyclelot":
            pass
        elif object_type not in KNOWN_USER_DEFINED_TYPES:
            results.append(_make_result(proxy, "ALERT", f"ObjectType '{getattr(proxy, 'ObjectType', None)}' is missing or not in the known user-defined type list."))
            continue
        else:
            continue

        bay_usage = _get_property_text(proxy, "PSet_InHouse_Motorcycle_Props", "BayUsage", _norm_upper)
        if bay_usage is None:
            results.append(_make_result(proxy, "ALERT", "PSet_InHouse_Motorcycle_Props.BayUsage is missing."))
            continue
        if bay_usage not in ALLOWED_BAY_USAGE:
            results.append(_make_result(proxy, "ALERT", f"BayUsage '{bay_usage}' is invalid."))
            continue

        if storey is None:
            results.append(_make_result(proxy, "ALERT", "Hosting storey could not be resolved from spatial structure."))
            continue

        lot_geom = _geometry_cache_entry(proxy, settings, geometry_cache)
        if not lot_geom["ok"]:
            results.append(_make_result(proxy, "ALERT", f"Motorcycle lot geometry is unusable: {lot_geom['reason']}."))
            continue

        candidate_spaces = spaces_by_storey.get(storey.id(), [])
        containing_spaces = []
        unreliable_space_geometry = False
        for space in candidate_spaces:
            space_geom = _geometry_cache_entry(space, settings, geometry_cache)
            if not space_geom["ok"]:
                unreliable_space_geometry = True
                continue
            try:
                contains_2d = space_geom["footprint"].buffer(CONTAIN_2D_TOL).covers(lot_geom["footprint"])
            except Exception:
                unreliable_space_geometry = True
                continue
            vertical_ok = (
                lot_geom["min_z"] >= space_geom["min_z"] - VERT_TOL
                and lot_geom["max_z"] <= space_geom["max_z"] + VERT_TOL
            )
            if contains_2d and vertical_ok:
                containing_spaces.append(space)

        if unreliable_space_geometry:
            results.append(_make_result(proxy, "ALERT", "Containing-space search is unreliable because one or more same-storey spaces have unusable geometry."))
            continue
        if len(containing_spaces) != 1:
            if len(containing_spaces) == 0:
                reason = "No unique same-storey IfcSpace fully contains the motorcycle lot geometry."
            else:
                reason = "More than one same-storey IfcSpace fully contains the motorcycle lot geometry."
            results.append(_make_result(proxy, "ALERT", reason))
            continue

        containing_space = containing_spaces[0]
        parking_environment = _get_property_text(
            containing_space,
            "PSet_InHouse_Space_Props",
            "ParkingEnvironment",
            _norm_upper,
        )
        if parking_environment is None:
            results.append(_make_result(proxy, "ALERT", "PSet_InHouse_Space_Props.ParkingEnvironment is missing on the containing space."))
            continue
        if parking_environment not in ALLOWED_PARKING_ENVIRONMENT:
            results.append(_make_result(proxy, "ALERT", f"ParkingEnvironment '{parking_environment}' is invalid."))
            continue

        if bay_usage in {"TANDEM", "EV", "LOADING"}:
            results.append(_make_result(proxy, "NA", f"BayUsage is {bay_usage}, which is exempt from this dimension check."))
            continue
        if parking_environment == "COVERED":
            results.append(_make_result(proxy, "NA", "Containing space ParkingEnvironment is COVERED, so the requirement does not apply."))
            continue
        if parking_environment == "SEMI":
            results.append(_make_result(proxy, "MANUAL_CHECK", "Containing space ParkingEnvironment is SEMI, so manual review is required."))
            continue

        storey_kind, storey_name = _storey_classification(storey)
        if storey_kind is None:
            results.append(_make_result(proxy, "ALERT", f"Storey name '{getattr(storey, 'Name', None)}' is not a valid level or basement name."))
            continue
        if storey_kind == "level":
            results.append(_make_result(proxy, "NA", f"Storey name '{storey_name}' is a level at 1 or above, so the requirement does not apply."))
            continue

        dims = _measure_length_width(lot_geom["footprint"])
        if dims is None:
            results.append(_make_result(proxy, "ALERT", "Motorcycle lot plan geometry is degenerate, so length and width could not be measured."))
            continue
        length, width = dims

        if length >= 10.0 - DIM_TOL or width >= 10.0 - DIM_TOL:
            results.append(
                _make_result(
                    proxy,
                    "MANUAL_CHECK",
                    f"Measured dimensions are width {width:.3f} m and length {length:.3f} m; at least one dimension is 10 m or greater.",
                )
            )
            continue

        if width >= 1.2 - DIM_TOL and length >= 2.5 - DIM_TOL:
            results.append(
                _make_result(
                    proxy,
                    "PASS",
                    f"Measured dimensions are width {width:.3f} m and length {length:.3f} m in OPEN basement parking, meeting the minimum size.",
                )
            )
        else:
            results.append(
                _make_result(
                    proxy,
                    "FAIL",
                    f"Measured dimensions are width {width:.3f} m and length {length:.3f} m, below the required 1.2 m by 2.5 m minimum.",
                )
            )
    # end compliance code
    return results
\end{lstlisting}

%% file: parts/annexF.tex
\onecolumn
\section{Regulatory Requirements Dataset (R1--R10)}
\label{annex:dataset-requirements}

Table~\ref{tab:dataset-requirements-1} and Table~\ref{tab:dataset-requirements-2} describe each of the ten hypothetical regulatory requirements introduced in Section~\ref{sec:dataset}: what the requirement is designed to stress-test in a synthesized checker, the out-of-distribution (OOD) scenario built into its test pair $D_{test}$, and the resulting train/test element counts.

\begin{table}[htbp]
\centering
\scriptsize
\setlength{\tabcolsep}{3pt}
\renewcommand{\arraystretch}{1.0}
\caption{Dataset requirements R1--R5: what each requirement tests, its held-out OOD scenario, and train/test set sizes (number of elements in $m_{train}$ / $m_{test}$).}
\label{tab:dataset-requirements-1}
\begin{tabular}{p{0.04\textwidth}p{0.37\textwidth}p{0.42\textwidth}p{0.05\textwidth}p{0.05\textwidth}}
\toprule
\textbf{Req.} & \textbf{Description} & \textbf{Test OOD Description} & \textbf{Train} & \textbf{Test} \\
\midrule

\textbf{R1} &
This requirement tests for ability to create robust code for measuring single rectangular element dimensions combined with standard flow logic. &
A lot on a $25^\circ$ sloped floor slab with its length running up the slope: the true 2.5~m $\times$ 5.05~m bay $\mathrm{PASS}$es, but the $x$-$y$ projection of its length ($\sim$4.6~m) falls below the 4.8~m minimum, so a checker measuring the plan projection rather than the lot's true (along-slope) dimensions wrongly $\mathrm{FAIL}$s. &
26 & 27 \\
\addlinespace

\textbf{R2} &
This requirement tests for ability to create code for spatial analysis between multiple elements combined with standard flow logic. &
Three lot-pairs share one aisle with identical back walls, but lot length varies per pair so the front clearance differs even though the back-wall spacing does not: this defeats a checker that infers one fixed aisle width for the whole row instead of measuring each lot's actual clearance. &
45 & 51 \\
\addlinespace

\textbf{R3} &
This requirement tests for ability to create code that validates contextual metadata and spatial relationships (building address, property sets, containing space, storey) before measuring motorcycle lot dimensions, combined with standard flow logic. &
Two U-shaped space probes that wrap only the ends of a parking lot, leaving its middle over an open notch: one lying flat in plan (defeats axis-aligned bounding-box containment), one standing upright so the lot's footprint projects onto solid space (defeats footprint/2D-overlap containment too). &
48 & 51 \\
\addlinespace

\textbf{R4} &
This requirement tests for ability to create code that infers a sewer line's connectivity from geometry alone (no \texttt{IfcRelConnects}/port relationships), distinguishing cylindrical from non-cylindrical pipes, finding manhole pairs joined by touching/intersecting pipe runs, and measuring centre-to-centre spacing, combined with standard flow logic. &
A manhole connected only to a pipe loop that exits one side and re-enters the same manhole on another side: there is no second manhole anywhere on the loop, so it cannot form a sewer line, but a checker that builds a generic connectivity graph without checking the two endpoints are distinct chambers would wrongly pair the manhole with itself. &
22 & 23 \\
\addlinespace

\textbf{R5} &
This requirement tests for ability to create code that measures staircase clear width within a vertical headroom band ($\geq$0.9~m up to 2.1~m above the stair surface), reducing realistic railing assemblies to their governing faces, accounting for the listed obstacle types (railings, columns, beams, stair flights, walls), and resolving a rich set of branches (mechanical-space / level-5+ / $\leq$2-tread exemptions, odd-shaped or curved flights, and missing/invalid tread counts), combined with standard flow logic. &
Two unseen geometries, each shown as a $\mathrm{PASS}$ and a $\mathrm{FAIL}$: (a) straight stairs whose runs are rotated off the $x$-$y$ axes (built from a fresh solid stepped-block geometry), so a checker that measures an axis-aligned bounding box reads the rotated diagonal instead of the true cross-run width; (b) a stair whose left and right rails are physically one continuous ``U-turn'' railing that runs up one edge, loops around the top landing and returns down the other, defeating a checker that assumes one distinct railing per side instead of measuring both governing faces of the single railing. &
12 & 16 \\
\bottomrule
\end{tabular}
\end{table}

\begin{table}[htbp]
\centering
\scriptsize
\setlength{\tabcolsep}{3pt}
\renewcommand{\arraystretch}{1.0}
\caption{Dataset requirements R6--R10: what each requirement tests, its held-out OOD scenario, and train/test set sizes (number of elements in $m_{train}$ / $m_{test}$).}
\label{tab:dataset-requirements-2}
\begin{tabular}{p{0.04\textwidth}p{0.37\textwidth}p{0.42\textwidth}p{0.05\textwidth}p{0.05\textwidth}}
\toprule
\textbf{Req.} & \textbf{Description} & \textbf{Test OOD Description} & \textbf{Train} & \textbf{Test} \\
\midrule

\textbf{R6} &
This requirement tests for ability to create code that orients real vendor-authored faceted furniture meshes (reused from the IFCNet dataset) to recover a sofa's back vs.\ seating long side, then measures the back side's flushness ($\leq$10~mm) and parallelism ($\leq$5$^\circ$) to a wall while confirming the opposite seating side is left open (no obstruction within 1~m), combined with classification of ambiguous sofa-named \texttt{USERDEFINED} furniture and a near-square-footprint $\mathrm{MANUAL\_CHECK}$ branch, and standard flow logic. &
\textit{NIL: only parametric variations, no OOD scenario.} &
10 & 10 \\
\addlinespace

\textbf{R7} &
This requirement tests for ability to create code that defines a 3~m-radius vertical tree protection zone on each outdoor \texttt{IfcGeographicElement}\slash\texttt{TREE} trunk axis and checks listed obstacle solids (walls, columns, slabs, stairs, other trees, distribution chambers, \ldots) for penetration into the cylinder interior, combined with classification of invalid user-defined types, storey-level exemptions (level 2+), and standard flow logic. Parametric tree canopies (conical through irregular non-uniform crowns) and a glass-house pavilion with landscape scenery exercise a realistic outdoor site. &
An isolated shrub with an \texttt{IfcDistributionChamberElement} whose face stands 2.0~m from the trunk axis ($<$3~m): defeats a checker that omits distribution chambers from the obstacle list and would wrongly $\mathrm{PASS}$. &
11 & 12 \\
\addlinespace

\textbf{R8} &
This requirement tests for ability to create code that exempts sufficiently-externalized planter boxes (\texttt{IfcBuildingElementProxy}\slash\texttt{planterbox}) from area computation, measuring the longest plan dimension ($\leq$1~m) and inscribed-circle thickness ($\leq$0.5~m), then resolving externalization positionally (on a \texttt{ROOF} slab, inside / straddling / outside the \texttt{IsExternal} facade footprint, or covered by a roof overhead), combined with standard flow logic. Two connected glass-house pavilions, a glazed level-2 bridge, a roof garden, balcony and shed planters exercise the $\mathrm{PASS}$ / $\mathrm{FAIL}$ / $\mathrm{MANUAL\_CHECK}$ branches. &
Three OOD scenarios: (a) a valid-size planter with a hexagonal footprint on external ground, open to the sky (a novel convex shape used nowhere in train); (b) a \texttt{USERDEFINED} proxy whose ObjectType is \texttt{flower\_trough} (not a standard type) $\rightarrow$ R8 step 2 \textbf{ALERT}; (c) a valid-size planter on the bridge deck, outside both facades but covered by the bridge roof slab overhead $\rightarrow$ $\mathrm{MANUAL\_CHECK}$, defeating a checker that treats any external deck planter as cleanly $\mathrm{PASS}$. &
27 & 30 \\
\addlinespace

\textbf{R9} &
This requirement tests for ability to create code that measures the clear height of doors (\texttt{IfcDoor}), analysing part structure to exclude rigid top/bottom frames where present, and checks it against the 2~m minimum, combined with standard flow logic for non-constant clear height inside the frame. Anchors reuse real vendor-authored single-element door geometries (from the IFCNet dataset), embedded with their Body representation decomposed one tessellated solid per source item; a synthetic framed door exercises the case where overall height passes but clear opening does not. &
Two unseen geometries: (a) arched glass door whose full-width springline clear opening is $\sim$1.84~m but whose arch peak reaches $\sim$2.28~m: defeats a checker that picks one convention for arched clear height instead of returning \textbf{MANUAL\_CHECK}; (b) carved timber door with relief ribs whose leaf clear height is 2.08~m: defeats a checker that mistakes protruding rib geometry for a reduced vertical opening and wrongly $\mathrm{FAIL}$s. &
5 & 7 \\
\addlinespace

\textbf{R10} &
This requirement tests for ability to create code that validates fire-escape connectivity for \texttt{IfcSpace} rooms, resolving labels from \texttt{InHouse\_Set}\slash\texttt{SpaceName}, identifying Egress Indicator Boxes on doors to fix permitted egress direction, treating compound touching spaces as one logical room, applying kitchen exemptions, and searching bounded escape routes to \texttt{Common Corridor} ($\leq$2 egress doors, $\leq$1 intermediate room) and from corridor to \texttt{Escape Stair Zone} (exactly 1 door), combined with connectivity/metadata \textbf{ALERT} branches and standard flow logic. &
Two OOD scenarios: (a) a room with valid egress to either of two separate \texttt{Common Corridor} spaces: tests target-set search rather than a single named destination; (b) a door carrying two EIBs (ambiguous egress direction) is excluded from route analysis, leaving its adjacent room with no usable egress: defeats a checker that picks an arbitrary direction or double-counts the door. &
39 & 43 \\
\bottomrule
\end{tabular}
\end{table}

%% file: parts/annexG.tex
\onecolumn
\section{Per-Scenario Difficulty, Cost, and Model-Choice Analysis}
\label{annex:per-scenario-analysis}

The aggregate results of Section~\ref{sec:experiments} pool over the ten requirements, but the requirements are far from homogeneous: some are intrinsically harder than others, some consume disproportionate token budgets, and for some the choice of backbone model materially changes the outcome. This annex breaks the results down per requirement (see Annex~\ref{annex:dataset-requirements} for what each requirement tests).

\subsection{Not All Scenarios Are Equally Hard}

\begin{figure}[H]
\centering
\includegraphics[width=0.5\textwidth]{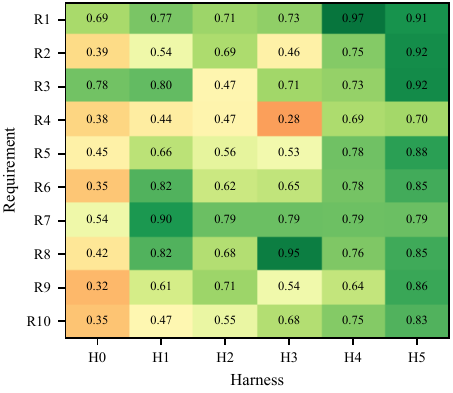}
\caption{Test union accuracy per requirement and harness, averaged over the four backbone models. Requirements demanding multi-element relational reasoning (R4, R5, R10) are consistently harder than per-element measurement requirements (R1, R3), but ARCHER (H5) compresses the difficulty spread, lifting every requirement to a cross-model mean of at least 0.70.}
\label{fig:scenario-difficulty}
\end{figure}

Figure~\ref{fig:scenario-difficulty} shows union accuracy per requirement and harness, averaged over the four backbone models. Averaged across all 24 (harness, model) configurations, the requirements span a wide difficulty range: from R1, the simplest single-element dimensional check (mean $A_{\cup}$ = 0.80), down to R4, geometric inference of sewer-line connectivity (0.49), with R10, the fire-escape route-search requirement, close behind (0.60). Difficulty tracks the nature of the computation: requirements solvable by per-element measurement with rich metadata (R1, R3, R7, R8) start high even under weak harnesses, whereas requirements demanding relational or topological reasoning across multiple elements, such as pairing manholes through touching pipe runs (R4), measuring clearance bands across railing assemblies (R5), or searching bounded egress routes through a space-adjacency graph (R10), remain difficult until the multi-agent harnesses are applied. The harness does not erase this ordering, but it compresses it: under ARCHER (H5) every requirement reaches a cross-model mean of at least 0.70, and the hardest requirements gain the most (R10: 0.35 $\rightarrow$ 0.83; R2: 0.39 $\rightarrow$ 0.92).

\subsection{Hard Scenarios Are Disproportionately Expensive}

\begin{figure}[H]
\centering
\includegraphics[width=\textwidth]{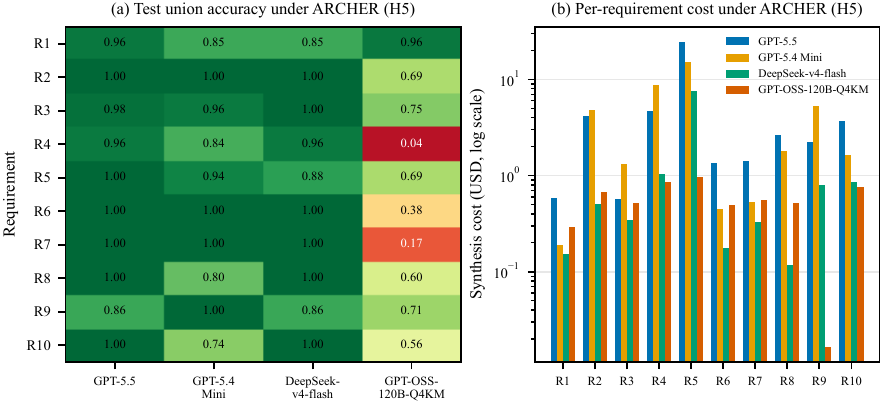}
\caption{Per-requirement results under ARCHER (Harness 5). (a) Test union accuracy by backbone model. (b) Synthesis cost per requirement (USD, log scale). Hard requirements (R4, R5) cost an order of magnitude more than easy ones for every backbone, and the on-premise model's failures concentrate on the relational-reasoning requirements.}
\label{fig:scenario-h5}
\end{figure}

Difficulty shows up in the token bill before it shows up in the accuracy column. Figure~\ref{fig:scenario-h5}(b) plots the synthesis cost of each requirement under ARCHER: because the evaluator keeps re-invoking the generator until train accuracy converges or the iteration cap is reached, hard requirements consume many more refinement rounds. The spread is large: for \texttt{GPT-5.5} the cheapest requirement (R3, \$0.57) and the most expensive (R5, \$24.35) differ by a factor of 42, and R5 alone accounts for 53\% of \texttt{GPT-5.5}'s and 64\% of \texttt{DeepSeek-v4-flash}'s total spend across all ten requirements. Per-requirement cost is thus a useful \emph{difficulty signal in itself}: a requirement that burns an outsized share of the budget without converging (e.g.\ \texttt{GPT-OSS-120B-Q4KM} spending \$0.85 on R4 yet finishing at 0.04) is a strong indicator that the configuration is under-powered for that scenario and the requirement should be escalated, whether to a stronger backbone or to human review.

\begin{table}[H]
\centering
\footnotesize
\setlength{\tabcolsep}{5pt}
\renewcommand{\arraystretch}{1.2}
\caption{Test union accuracy and synthesis cost (USD, in parentheses) per requirement under ARCHER (Harness 5). \textbf{Bold} marks the best accuracy per requirement; when tied, all tied entries are bold.}
\label{tab:per-scenario-h5}
\begin{tabular}{@{}l cccc@{}}
\toprule
\textbf{Req.} & \textbf{GPT-5.5} & \textbf{GPT-5.4 Mini} & \textbf{DeepSeek-v4-flash} & \textbf{GPT-OSS-120B-Q4KM} \\
\midrule
R1 & \textbf{0.963} (\$0.59) & 0.852 (\$0.19) & 0.852 (\$0.15) & \textbf{0.963} (\$0.29) \\
R2 & \textbf{1.000} (\$4.20) & \textbf{1.000} (\$4.79) & \textbf{1.000} (\$0.51) & 0.686 (\$0.67) \\
R3 & 0.980 (\$0.57) & 0.961 (\$1.31) & \textbf{1.000} (\$0.35) & 0.750 (\$0.51) \\
R4 & \textbf{0.958} (\$4.74) & 0.840 (\$8.68) & \textbf{0.958} (\$1.04) & 0.042 (\$0.85) \\
R5 & \textbf{1.000} (\$24.35) & 0.938 (\$15.23) & 0.875 (\$7.65) & 0.688 (\$0.97) \\
R6 & \textbf{1.000} (\$1.35) & \textbf{1.000} (\$0.45) & \textbf{1.000} (\$0.18) & 0.385 (\$0.49) \\
R7 & \textbf{1.000} (\$1.42) & \textbf{1.000} (\$0.53) & \textbf{1.000} (\$0.33) & 0.167 (\$0.55) \\
R8 & \textbf{1.000} (\$2.66) & 0.800 (\$1.82) & \textbf{1.000} (\$0.12) & 0.600 (\$0.52) \\
R9 & 0.857 (\$2.24) & \textbf{1.000} (\$5.35) & 0.857 (\$0.80) & 0.714 (\$0.02) \\
R10 & \textbf{1.000} (\$3.74) & 0.744 (\$1.66) & \textbf{1.000} (\$0.85) & 0.562 (\$0.76) \\
\bottomrule
\end{tabular}
\end{table}

\subsection{When Does a Stronger (or More Expensive) Model Make Sense?}

Table~\ref{tab:per-scenario-h5} and Figure~\ref{fig:scenario-h5}(a) support a per-scenario reading of the deployment-tier question from Section~\ref{sec:cost-analysis}.

\textbf{For the majority of requirements, the self-hosted tier is already sufficient.} \texttt{DeepSeek-v4-flash} matches or beats the frontier model on seven of ten requirements (perfect 1.000 on R2, R3, R6, R7, R8, R10, and tied 0.958 on R4), typically at a tenth of the cost. An organization restricted to its own cloud loses essentially nothing on these scenarios.

\textbf{Escalating to the frontier tier pays off precisely on the hardest scenarios.} \texttt{GPT-5.5}'s advantage concentrates on R5 (1.000 vs.\ 0.875 for \texttt{DeepSeek-v4-flash}), the staircase clear-width requirement whose rotated-geometry and continuous-railing test cases defeat bounding-box shortcuts, and it is also the only backbone to fully solve R10's route-search logic alongside \texttt{DeepSeek-v4-flash}. Notably, this peak accuracy is bought with the single most expensive synthesis in the study (\$24.35 for R5): the frontier model does not make hard scenarios cheap, it makes them \emph{solvable}.

\textbf{Backbone weaknesses are scenario-specific, not uniform.} \texttt{GPT-5.4 Mini} solves R9 perfectly (the only backbone to do so) yet is the weakest API model on R10 (0.744); the on-premise \texttt{GPT-OSS-120B-Q4KM} nearly matches the frontier model on the metadata-flow requirements R1 (0.963) and R9 (0.714 at \$0.02) but collapses on relational geometry (R4: 0.042; R7: 0.167). Averaged scores thus understate both how useful cheap models are on easy scenarios and how unsuitable they are for hard ones.

\textbf{Implication: route requirements to tiers.} Because ARCHER scores every checker against $D_{train}$, per-requirement accuracy and cost are observable at synthesis time, making tiered routing practical: synthesize on the cheapest deployable tier first, escalating only when train-time accuracy stalls or the refinement loop overruns its token budget. On our dataset, an oracle router picking the cheapest best-accuracy Harness-5 configuration per requirement would reach a mean union accuracy of 0.992 at \$33.37 total, beating the all-frontier column (0.976, \$45.86) at 27\% lower cost, with eight of ten requirements never leaving the self-hostable tiers.